\documentclass[11pt]{article}
\usepackage{graphicx}
\usepackage{indentfirst,csquotes}
\usepackage{booktabs}

\usepackage{amssymb,amsthm,amsmath}
\usepackage{xcolor,paralist,hyperref,titlesec,fancyhdr,etoolbox}

\hypersetup{ colorlinks=true, linkcolor=black, filecolor=black, urlcolor=black }

\usepackage{titletoc}
\usepackage[resetlabels]{multibib}

\newcites{Paper}{References}
\newcites{Supp}{Supplementary References}

\usepackage{caption} 
\usepackage[a4paper, top=2.5cm, bottom=2.5cm, left=2.5cm, right=2.5cm]{geometry}

\hypersetup{
pdftitle={General linear hypothesis testing in ill-conditioned functional response model},
pdfsubject={stat},
pdfauthor={Łukasz Smaga, Natalia Stefańska},
pdfkeywords={bootstrap methods, functional data analysis, functional regression, functional response model, general hypothesis testing, ill-conditioned design},
}

\begin{document}

\title{\textbf{General linear hypothesis testing\\ in ill-conditioned functional response model}}

\date{} 					

\author{Łukasz Smaga$^1$ \and Natalia Stefańska$^1$\thanks{Corresponding author\\ Email address: ls@amu.edu.pl, natste16@st.amu.edu.pl (Łukasz Smaga, Natalia Stefańska)}}

\maketitle
\begin{center}
\footnotesize\textsuperscript{1}Adam Mickiewicz University, Faculty of Mathematics and Computer Science, Poznań (Poland) \\
\end{center}

\begin{abstract}
The paper concerns inference in the ill-conditioned functional response model, which is a part of functional data analysis. In this regression model, the functional response is modeled using several independent scalar variables. To verify linear hypotheses, we develop new test statistics by aggregating pointwise statistics using either integral or supremum. The new tests are scale-invariant, in contrast to the existing ones. To construct tests, we use different bootstrap methods. The performance of the new tests is compared with the performance of known tests through a simulation study and an application to a real data example.\\

\end{abstract}

\startcontents[sections]

\section{Introduction}
Functional data analysis (FDA) is a specialized branch of statistics that involves analyzing data in the form of functions or curves. Due to improvements in measuring technology, discrete data are collected as measurements of some variable over time or space and then transformed into functional data. The functional representation of data is used in many disciplines - for example in meteorology, where temperature is observed over time. In the last 20 years, statistical methods, including classification, clustering, dimension reduction, regression, and statistical hypothesis testing, have been developed for functional data. For methodology and real data examples in FDA, we refer to the textbooks \citePaper{HorKok2012, ramsay_silverman_2005, Zhang2013}.

One of the functional regression models is the functional response model, where the response variable is functional, while the predictors are scalars~\citePaper{HorKok2012}. For the full-rank functional response model (where the design matrix is of full rank), Shen and Faraway~\citePaper{ShenFar1}, Zhang and Chen~\citePaper{zhangchen2007}, and Zhang~\citePaper{zhang11} proposed the $L^2$-norm-based and $F$-type tests. These test procedures were constructed under specific assumptions such as Gaussianity or large sample sizes. When those assumptions were not satisfied, Zhang~\citePaper{Zhang2013} proposed nonparametric bootstrap tests based on the test statistics of the aforementioned procedures. Moreover, Smaga~\citePaper{smaga2021} investigated the properties of two new tests for this problem - the globalizing pointwise $F$ test and the $F_{\max}$-test. The properties were examined in numerical studies, which suggested that the new methods achieved better outcomes compared to previously known test procedures.

In practice, however, the full-rank design matrix is not always the case (see, for example, Sections~\ref{sec_4}-\ref{sec_5}). In such cases, we have the ill-conditioned functional response model (ICFRM), which is a generalization of the full-rank model. This issue was addressed by Zhang~\citePaper{Zhang2013}, who developed three methodologies for finding and making inferences about estimable parametric functions. These methods are the generalized inverse method, the reparameterization method, and the side-condition method. Although developed from different statistical perspectives, they are fundamentally equivalent.  Zhang~\citePaper{Zhang2013} also adopted the $L^2$-norm-based and $F$-type tests for the ICFRM. However, their finite sample properties were not investigated. In this paper, we fill this gap, although this is not our main goal. We also noted that these tests are not scale-invariant in the sense of Guo, Zhou, and Zhang~\citePaper{GuoEtAl2019}. This can have serious consequences for their properties. Thus, we construct the globalizing pointwise $F$ test and the $F_{\max}$-test for the general linear hypothesis testing problem in the ICFRM, which are scale-invariant and more powerful than the tests in~\citePaper{Zhang2013}.

The structure of the remaining sections is as follows. In Section~\ref{sec_2}, the ICFRM and the general linear hypothesis testing problem are introduced. The previously established test statistics, the new test statistics, and the methodologies for creating new tests are presented in Section~\ref{sec_3}. In Sections~\ref{sec_4} and~\ref{sec_5}, we provide a simulation study and a real data example for the considered methods, respectively. They are based on the audible noise data presented in~\citePaper{Zhang2013}. Section~\ref{sec_6} offers concluding remarks. 

\section{Statistical hypotheses in ill-conditioned functional response model}
\label{sec_2}
Assume the functions $y_i(t)$, $t\in[a,b]$, $a,b\in\mathbb{R},a<b$ can be represented by a functional response model $$y_i(t) = \mathbf{x}_i^{\top} \boldsymbol{\beta}(t) + v_i(t),$$ where $i = 1,\dots,n$, $$\mathbf{x}_i = (1, x_{i1}, \dots, x_{ip})^{\top}$$ are $(p+1)$-dimensional predictors, $$\boldsymbol{\beta}(t)=(\beta_0(t), \beta_1(t), \dots, \beta_p(t))^{\top}$$ is a $(p+1)$-dimensional vector of unknown coefficient functions, and $v_i(t)$ are independent stochastic processes with a common distribution - with zero mean and a covariance function $\gamma(s,t)$, $s,t\in[a,b]$. Note that we do not assume any specific distribution. This model can be expressed in the matrix form $$\mathbf{y}(t) = \mathbf{X} \boldsymbol{\beta}(t) + \mathbf{v}(t),\ t\in[a,b],$$ where $$\mathbf{y}(t) = (y_1(t), \dots, y_n(t))^{\top},\ \mathbf{X} = (\mathbf{x}_1, \dots, \mathbf{x}_n)^{\top},\ \mathbf{v}(t) = (v_1(t), \dots, v_n(t))^{\top}.$$

In this model, we assume that the matrix $\mathbf{X}$ is not full-rank, i.e., $\mathbf{X}^{\top}\mathbf{X}$ is not invertible, and let $rank(\mathbf{X})=k<p+1<n$. This type of model is called the ill-conditioned functional response model (ICFRM, Zhang~\citePaper{Zhang2013}, Chapter~7). The estimation of the vector $\boldsymbol{\beta}$ of unknown parameters is important and challenging. For this problem, three equivalent methods were proposed in~\citePaper{Zhang2013}. Here, we will use the generalized inverse method, which results in the estimator $$\hat{\boldsymbol{\beta}}(t) = (\mathbf{X}^{\top}\mathbf{X})^-\mathbf{X}^{\top}\mathbf{y}(t),$$ $t\in[a,b]$, where $(\mathbf{X}^{\top}\mathbf{X})^-$ is a generalized inverse of $\mathbf{X}^{\top}\mathbf{X}$. However, $\hat{\boldsymbol{\beta}}(t)$ is biased for $\boldsymbol{\beta}(t)$. Thus, $\boldsymbol{\beta}(t)$ itself is not estimable. In general, the linear parametric function $\mathbf{C}\boldsymbol{\beta}(t)$ is called estimable if it has an unbiased estimator that is a linear function of $\mathbf{y}(t)$, where $\mathbf{C}$ is $q\times(p+1)$ constant and known matrix. Theorem~7.3 in~\citePaper{Zhang2013} presents the condition under which $\mathbf{C}\boldsymbol{\beta}(t)$ is estimable. In such a case, $\mathbf{C}\hat{\boldsymbol{\beta}}(t)$ is the unbiased estimator for $\mathbf{C}\boldsymbol{\beta}(t)$.

It is of interest to solve the general linear hypothesis testing (GLHT) problem:
\begin{align}\label{GLHT}
H_0:\mathbf{C}\boldsymbol{\beta}(t)=\mathbf{c}(t)\ \forall t\in [a,b]\text{ vs. } H_1: \mathbf{C}\boldsymbol{\beta}(t) \neq \mathbf{c}(t) \text{ for some }t \in [a,b],
\end{align}
where $\mathbf{C}$ is a full-rank matrix with rank $q \leq k < p+1 < n$, and $\mathbf{c}(t)=(c_1(t), \dots, c_q(t))^{\top}$ is $q \times 1$ vector of known functions. Usually, $\mathbf{c}(t)$ is a zero vector. When $\mathbf{C}\boldsymbol{\beta}(t)$ is estimable, then the GLHT is called testable.

\section{Test procedures}
\label{sec_3}
To verify~\eqref{GLHT}, Zhang~\citePaper{Zhang2013} used the $L^2$-norm-based tests and $F$-type tests based on the following test statistics:
\begin{align*}
T_n = \int_a^b SSH_n(t) \ dt,\ F_n = \frac{\int_a^b SSH_n(t) \ dt/q}{\int_a^b SSE_n(t) \ dt / (n-k)},
\end{align*}
respectively, where $$SSH_n(t) = (\mathbf{C}\hat{\boldsymbol{\beta}}(t) - \mathbf{c}(t))^{\top} (\mathbf{C}(\mathbf{X^{\top}X})^-\mathbf{C}^{\top})^{-1} (\mathbf{C}\hat{\boldsymbol{\beta}}(t) - \mathbf{c}(t))$$ is the pointwise sum of squares for hypothesis, $$SSE_n(t) = (n-k)\hat{\gamma}(t,t)$$ is the pointwise sum of squares for error, and $$\hat{\gamma}(s,t)=(n-k)^{-1}\mathbf{y}(s)^{\top}(\mathbf{I}_n-\mathbf{X}(\mathbf{X}^{\top}\mathbf{X})^-\mathbf{X}^{\top})\mathbf{y}(t)$$ is an unbiased estimator of the covariance function $\gamma(s,t)$. Note, however, that $T_n$ is based only on the information from $SSH_n(t)$, while $F_n$ uses the information from both $SSH_n(t)$ and $SSE_n(t)$ but in a separate way. To find better tests, we follow the idea of Smaga~\citePaper{smaga2021}, who introduced the following test statistics:
$$G_n = \frac{1}{q}\int_a^b\frac{SSH_n(t)}{\hat{\gamma}(t,t)} \ dt,\ F_{\max, n} = \frac{1}{q} \sup_{t \in [a,b]} \Bigg\{\ \frac{SSH_n(t)}{\hat{\gamma}(t,t)} \Bigg\}.$$

In Sections~\ref{sec_4} and \ref{sec_5}, we will observe their good finite sample properties. For now, however, we mention a property that distinguishes the new tests from the previous ones. The test statistics $G_n$ and $ F_{\max, n}$ are scale-invariant in the sense of \citePaper{GuoEtAl2019}, i.e., they do not change when the functional data in $\mathbf{y}(t)$ are multiplied by any fixed function $h:[a,b]\rightarrow\mathbb{R}$ with $h(t)\neq 0$ for all $t\in[a,b]$. We denote the scaled functional responses with a superscript $h$, i.e., $\mathbf{y}^h(t):=h(t)\mathbf{y}(t)$, $t\in[a,b]$. The corresponding null hypothesis is $H_0^h:\mathbf{C}\boldsymbol{\beta}^h(t)=\mathbf{c}^h(t)$ for all $t\in[a,b]$, where $\boldsymbol{\beta}^h(t):=h(t)\boldsymbol{\beta}(t)$ and $\mathbf{c}^h(t):=h(t)\mathbf{c}(t)$. Then, we have the scale-invariance of the pointwise test statistic
$$\frac{SSH_n^h(t)}{\hat{\gamma}^h(t,t)}=\frac{(\mathbf{C}\hat{\boldsymbol{\beta}}^h(t) - \mathbf{c}^h(t))^{\top} (\mathbf{C}(\mathbf{X^{\top}X})^-\mathbf{C}^{\top})^{-1} (\mathbf{C}\hat{\boldsymbol{\beta}}^h(t) - \mathbf{c}^h(t))}{(n-k)^{-1}\mathbf{y}^h(t)^{\top}(\mathbf{I}_n-\mathbf{X}(\mathbf{X}^{\top}\mathbf{X})^-\mathbf{X}^{\top})\mathbf{y}^h(t)}=\frac{SSH_n(t)}{\hat{\gamma}(t,t)}$$
for all $t\in[a,b]$, where $\hat{\boldsymbol{\beta}}^h(t)=h(t)\hat{\boldsymbol{\beta}}(t)$. Consequently, the $G_n$ and $F_{\max, n}$ statistics are also scale-invariant, while $T_n$ and $F_n$ are not. In the simulation study and real data example, we will examine the consequences of this.

To construct tests based on $G_n$ and $F_{\max,n}$, we use two bootstrap methods. We also investigated other methods similar to those in~\citePaper{smaga2021}, but the results were not satisfactory. The first bootstrap method is the nonparametric bootstrap ($G_n^{nb}$ and $F_{\max,n}^{nb}$ tests) used in~\citePaper{Zhang2013}. We randomly generate (with replacement) a large number $M$ of bootstrap samples $\hat{v}_i^{\star, m}(t)$, $t\in[a,b]$, $i=1,\dots,n$, $m=1,\dots,M$ from the estimated subject-effect functions $$\hat{v}_i(t) = y_i(t) - \mathbf{x}_i^{\top}\hat{\boldsymbol{\beta}}(t),\ t\in[a,b],\ i=1,\dots,n.$$ Additionally, we compute $$\hat{\boldsymbol{\beta}}^{\star, m}(t) = (\mathbf{X}^{\top}\mathbf{X})^-\mathbf{X}^{\top}\mathbf{y}^{\star, m}(t),$$ where $$\mathbf{y}^{\star, m}(t) = \mathbf{X}\hat{\boldsymbol{\beta}}(t) + \hat{\mathbf{v}}^{\star,m}$$ and $$\hat{\mathbf{v}}^{\star,m} = (\hat{v}_1^{\star, m}(t), \dots, \hat{v}_n^{\star, m}(t))^{\top}.$$ Next, we compute
\begin{align*}
SSH_n^{\star, m}(t) &= (\hat{\boldsymbol{\beta}}^{\star, m}(t) - \hat{\boldsymbol{\beta}}(t))^{\top}\mathbf{C}^{\top} (\mathbf{C}(\mathbf{X}^{\top}\mathbf{X})^-\mathbf{C}^{\top})^{-1}\mathbf{C}(\hat{\boldsymbol{\beta}}^{\star, m}(t) - \hat{\boldsymbol{\beta}}(t)),\\
SSE_n^{\star, m}(t) &= \mathbf{y}^{\star, m}(t)^{\top} (\mathbf{I}_n - \mathbf{X}(\mathbf{X}^{\top}\mathbf{X})^-\mathbf{X}^{\top})\mathbf{y}^{\star, m}(t).
\end{align*}
The $G_n$ and $F_{\max, n}$ nonparametric bootstrap test statistics are as follows:
$$G_n^{\star, m} = \frac{1}{q}\int_a^b\frac{SSH_n^{\star, m}(t)}{SSE_n^{\star, m}(t) / (n-k)} \ dt,\ F_{\max, n}^{\star, m} = \frac{1}{q} \sup_{t \in [a,b]} \Bigg\{\ \frac{SSH_n^{\star, m}(t)}{SSE_n^{\star, m}(t) / (n-k)} \Bigg\},$$
respectively, and $m = 1, \dots, M$. Finally, the $p$-values are $$M^{-1}\sum_{m=1}^MI(G_n^{\star, m}>G_n)$$ and $$M^{-1}\sum_{m=1}^MI(F_{\max,n}^{\star, m}>F_{\max,n}),$$ where $I(A)$ is the indicator function on the set~$A$.

The second bootstrap method is the parametric bootstrap ($G_n^{pb}$ and $F_{\max,n}^{pb}$ tests), which is different from that considered in~\citePaper{smaga2021}. To mimic the data under the null hypothesis, we generate the parametric bootstrap samples $\mathbf{y}^{*,m}(t)=(y_1^{*,m}(t),\dots,y_n^{*,m}(t))^{\top}$, where $y_1^{*,m},\dots,y_n^{*,m}$ are independent Gaussian processes with a zero mean function and a covariance function equal to $\hat{\gamma}$. For these samples, we calculate 
$$SSH_n^{*, m}(t) = (\mathbf{C}\hat{\boldsymbol{\beta}}^{*, m}(t))^{\top}(\mathbf{C}(\mathbf{X}^{\top}\mathbf{X})^-\mathbf{C}^{\top})^{-1}(\mathbf{C}\hat{\boldsymbol{\beta}}^{*, m}(t)),$$
where $$\hat{\boldsymbol{\beta}}^{*, m}(t)=(\mathbf{X}^{\top}\mathbf{X})^-\mathbf{X}^{\top}\mathbf{y}^{*,m}(t).$$ The rest of the procedure follows the same steps as for the nonparametric bootstrap, except that $\mathbf{y}^{\star,m}(t)$ is replaced with $\mathbf{y}^{*,m}(t)$.

\section{Simulation study}
\label{sec_4}
In this section, we study the finite sample properties (i.e., size control and power) of the new tests from Section~\ref{sec_3} and the tests in~\citePaper{Zhang2013} as competitors, i.e., the $L^2$-norm-based and $F$-type tests using both naive ($T_n^N$ and $F_n^N$) and bias-reduced ($T_n^B$ and $F_n^B$) estimation methods, as well as the nonparametric bootstrap method ($T_n^{nb}$ and $F_n^{nb}$). Similarly to~\citePaper{zhang11}, the simulation study is based on a real data set. However, for the ICFRM, we consider the audible noise data. This data set was collected during studies on reducing audible noise levels of alternators~\citePaper{nairetal2002}. It is available on the following website: \url{https://blog.nus.edu.sg/stazjt2020/research/monographs/analysis-of-variance-for-functional-data/}. 

Alternators generate audible noise during rotation. Due to recent technological improvements, engine noise has been significantly minimized, making the noise from alternators more evident and raising concerns about quality. To resolve this issue, an engineering team carried out a robust design study to measure the impact of the following seven assembly process factors on noise levels: ``Through Bolt Torque'' (A), ``Rotor Balance'' (B), ``Stator Varnish'' (C), ``Air Gap Variation'' (D), ``Stator Orientation'' (E), ``Housing Stator Slip Fit'' (F), and ``Shaft Radial Alignment'' (G). Among these, factor D is the noise factor, while the others are control factors. Each factor has two levels: low and high. The study measured audible noise levels at various rotating speeds. Microphones placed at different positions near the alternator captured the sound, which was then transformed into sound pressure levels. The study adopted a $2^{7-2}$ fractional factorial design involving seven factors at two levels each, resulting in a total of thirty-two runs for the experiment. Additionally, this design was supplemented with four additional replications at the high levels of all factors, resulting in a total of thirty-six runs for the experiment. For each response curve, forty-three measurements of sound pressure levels (in decibels) were taken, with rotating speeds ranging from 1,000 to 2,500 revolutions per minute.

\begin{figure}[t] 
    \centering
    \includegraphics[width=0.9\textwidth,height=0.33\textheight]{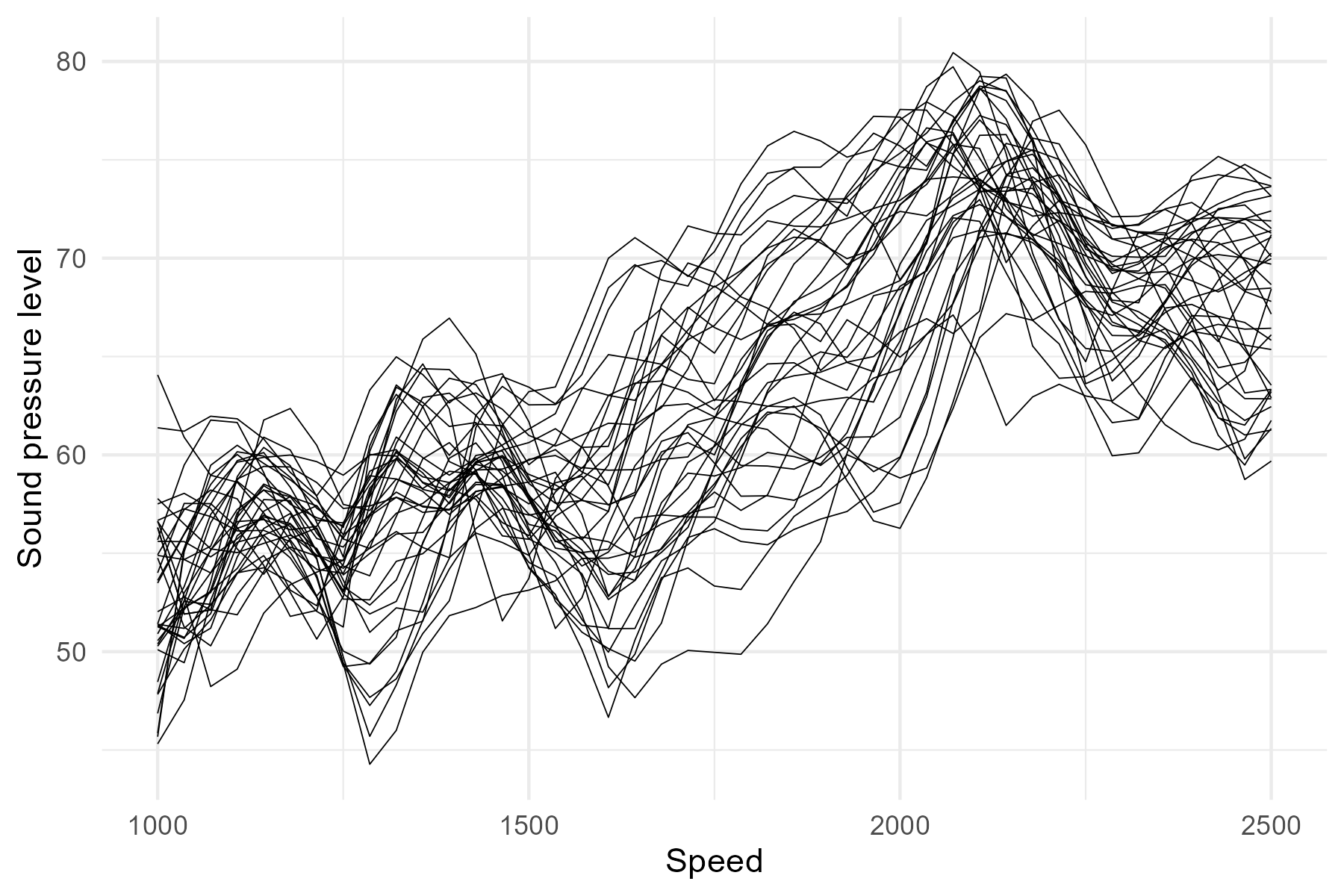}
\caption{Thirty-six sound pressure level curves.} 
\label{fig_1}
\end{figure}

In this experiment, we obtain 36 functional responses $y_1(t),\dots,y_{36}(t)$ measured at the 43 design time points (Figure~\ref{fig_1}). Each of these functional responses corresponds to an appropriate combination of factors A-G under the ICFRM with $$\boldsymbol{\beta}(t)=(\eta(t),\alpha_{11}(t),\alpha_{12}(t),\alpha_{21}(t),\alpha_{22}(t),\dots,\alpha_{71}(t),\alpha_{72}(t))^{\top},$$ where $\eta(t)$ is the grand mean function, and $\alpha_{ij}(t)$, $i=1,\dots,7$, $j=1,2$ are the main-effect functions of the factors. Namely, the matrix $\mathbf{X}$ presented in the supplement is of size $36\times 15$ and of rank $8$.

Following Zhang~\citePaper{zhang11}, we generate subject-effect functions $v_i(t)$, $t\in[0,1]$, $i=1,\dots,36$ using three different cases:

\textbf{Case 1}: Let $$v_i(t) = \sum_{s=1}^{m_0}\xi_{is}\psi_s(t),$$ where $m_0 \in \mathbb{N}$ is odd, $\xi_{is}$ are independent random variables of normal distribution $N(0, \lambda_s), \ \lambda_s = \rho^s, \ s = 1, \dots, m_0$ and $$\psi_1(t) = 1, \ \psi_{2r}(t) = \sqrt{2}\sin(2\pi rt),\ \text{and}\ \psi_{2r+1}(t) = \sqrt{2}\cos(2\pi rt)$$ are orthonormal basis functions, where $r = 1, \dots, (m_0 - 1)/2$. We set $m_0 = 13$ and $\rho = 0.1, 0.3, 0.5, 0.7, 0.9$.

\textbf{Case 2}: The only difference compared to Case~1 is that $\xi_{is} = \sqrt{\lambda_s}t_{is}/\sqrt{2}$, where $t_{is}$ are random variables of $t$-distribution with four degrees of freedom.

\textbf{Case 3}: The functions $v_i(t)$ are independent and identically distributed standard Wiener processes characterized by the dispersion parameter $h^2 = 0.3^2$.

Note the greater $\rho$ is the smaller the correlation between pointwise observations of functional data. For testing the null hypothesis~\eqref{GLHT}, we use the following matrix:
\begin{equation}
\label{mat_c}
    \mathbf{C} = \left(
\begin{array}{ccccccccccccccc}
   0 & 1 & -1 & 0 & 0 & 0 & 0 & 0 & 0 & 0 & 0 & 0 & 0 & 0 & 0 \\
   0 & 0 & 0 & 1 & -1 & 0 & 0 & 0 & 0 & 0 & 0 & 0 & 0 & 0 & 0 \\
   0 & 0 & 0 & 0 & 0 & 1 & -1 & 0 & 0 & 0 & 0 & 0 & 0 & 0 & 0 \\
   0 & 0 & 0 & 0 & 0 & 0 & 0 & 1 & -1 & 0 & 0 & 0 & 0 & 0 & 0 \\
   0 & 0 & 0 & 0 & 0 & 0 & 0 & 0 & 0 & 1 & -1 & 0 & 0 & 0 & 0 \\
   0 & 0 & 0 & 0 & 0 & 0 & 0 & 0 & 0 & 0 & 0 & 1 & -1 & 0 & 0 \\
   0 & 0 & 0 & 0 & 0 & 0 & 0 & 0 & 0 & 0 & 0 & 0 & 0 & 1 & -1 \\
\end{array}
\right)
\end{equation}
and $\mathbf{c}(t)=\mathbf{0}_7$ for $t\in[0,1]$. The matrix $\mathbf{C}$ corresponds to testing the main-effect contrast functions of the seven factors A-G, which are estimable.  To study the size and power of the tests, let $\boldsymbol{\beta} = \delta\hat{\boldsymbol{\beta}}$, where $\delta \geq 0$. When $\delta=0$, the null hypothesis is true and the empirical sizes of the tests are investigated. In contrast, $\delta \neq 0$ results in a false null hypothesis, thus leading to the study of the power of the tests. To show the importance of the scale-invariance of the new test, in contrast to those based on $T_n$ and $F_n$, we conduct the same simulations as above, but with the functional observations $y_i^h(t)=h(t)y_i(t)$, where $h(t)=1/(t+1/43)$ as in~\citePaper{GuoEtAl2019}. This case is referred to as ``with scaling''. We generate $1000$ simulation samples to evaluate the performance of the tests and employ $1000$ bootstrap samples for the bootstrap tests. We set a significance level $\alpha = 0.05$. Experiments were conducted in the R program~\citePaper{Rcore}. The code is available from the authors upon request.

\begin{figure}[!t] 
    \centering
    \includegraphics[width=0.99\textwidth,height=0.45\textheight]{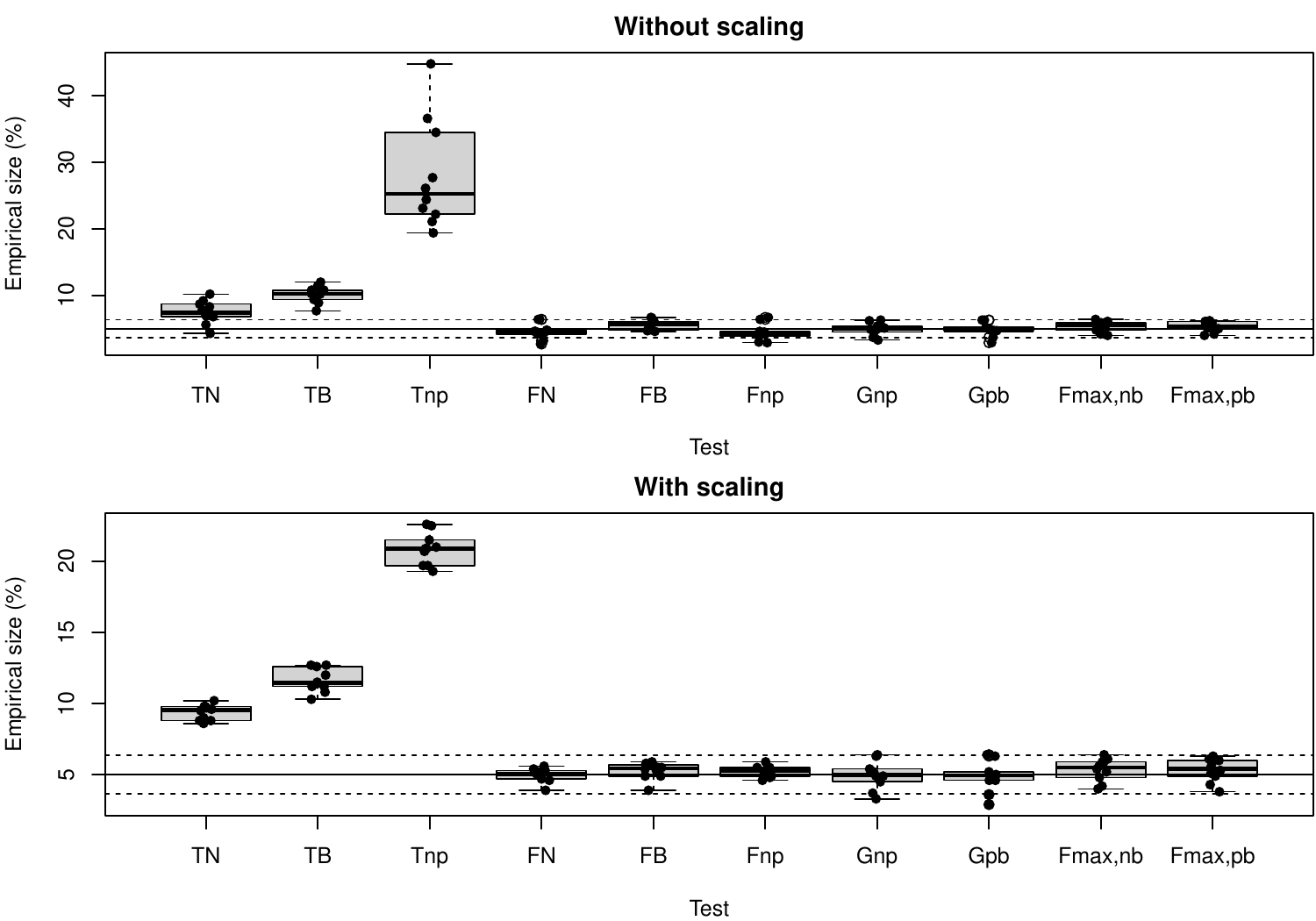}
\caption[Box-and-whisker plots for the empirical sizes obtained in Cases~1-2]{Box-and-whisker plots for the empirical sizes (as percentages) obtained in Cases~1-2.} 
\label{fig_s_1}
\end{figure}

\begin{figure}[!ht] 
    \centering
    \includegraphics[width=0.99\textwidth,height=0.9\textheight]{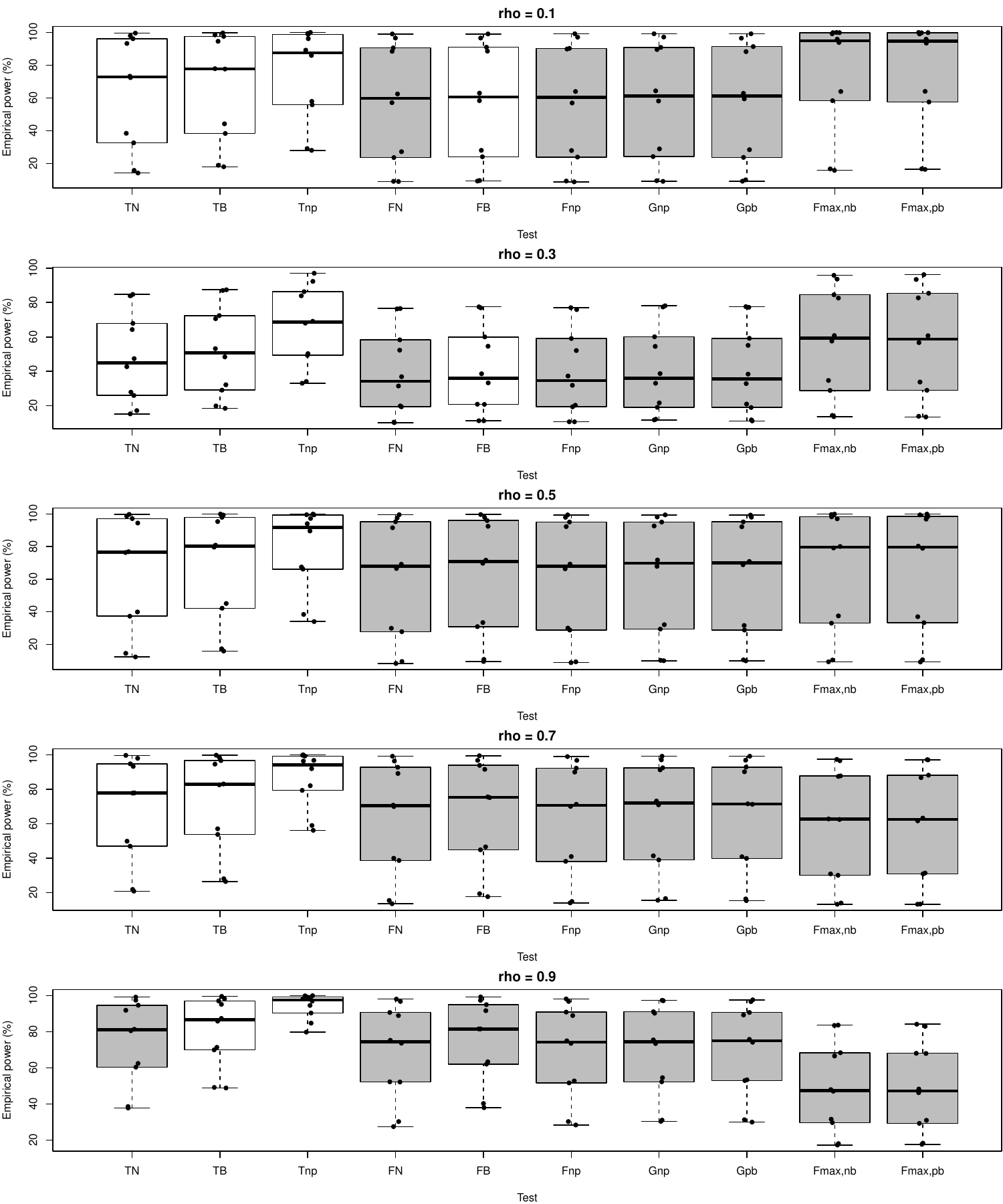}
\caption[Box-and-whisker plots for the empirical powers obtained in Cases~1-2 without scaling]{Box-and-whisker plots for the empirical powers (as percentages) obtained in Cases~1-2 without scaling. White boxplots represent too liberal tests.} 
\label{fig_s_2}
\end{figure}

\begin{figure}[!ht] 
    \centering
    \includegraphics[width=0.99\textwidth,height=0.9\textheight]{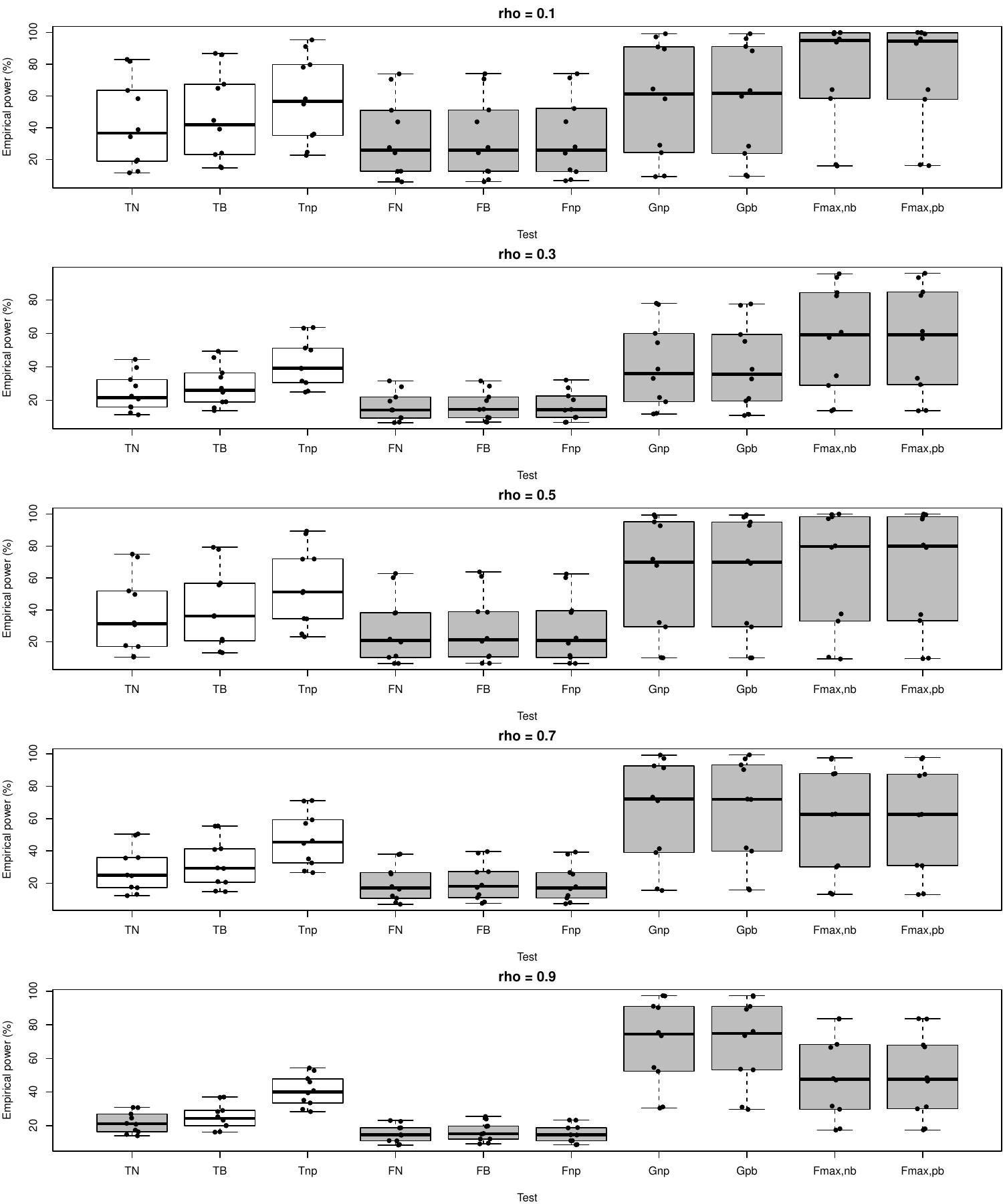}
\caption[Box-and-whisker plots for the empirical powers obtained in Cases~1-2 with scaling]{Box-and-whisker plots for the empirical powers (as percentages) obtained in Cases~1-2 with scaling. White boxplots represent too liberal tests.} 
\label{fig_s_3}
\end{figure}

\begin{figure}[!ht] 
    \centering
    \includegraphics[width=0.99\textwidth,height=0.45\textheight]{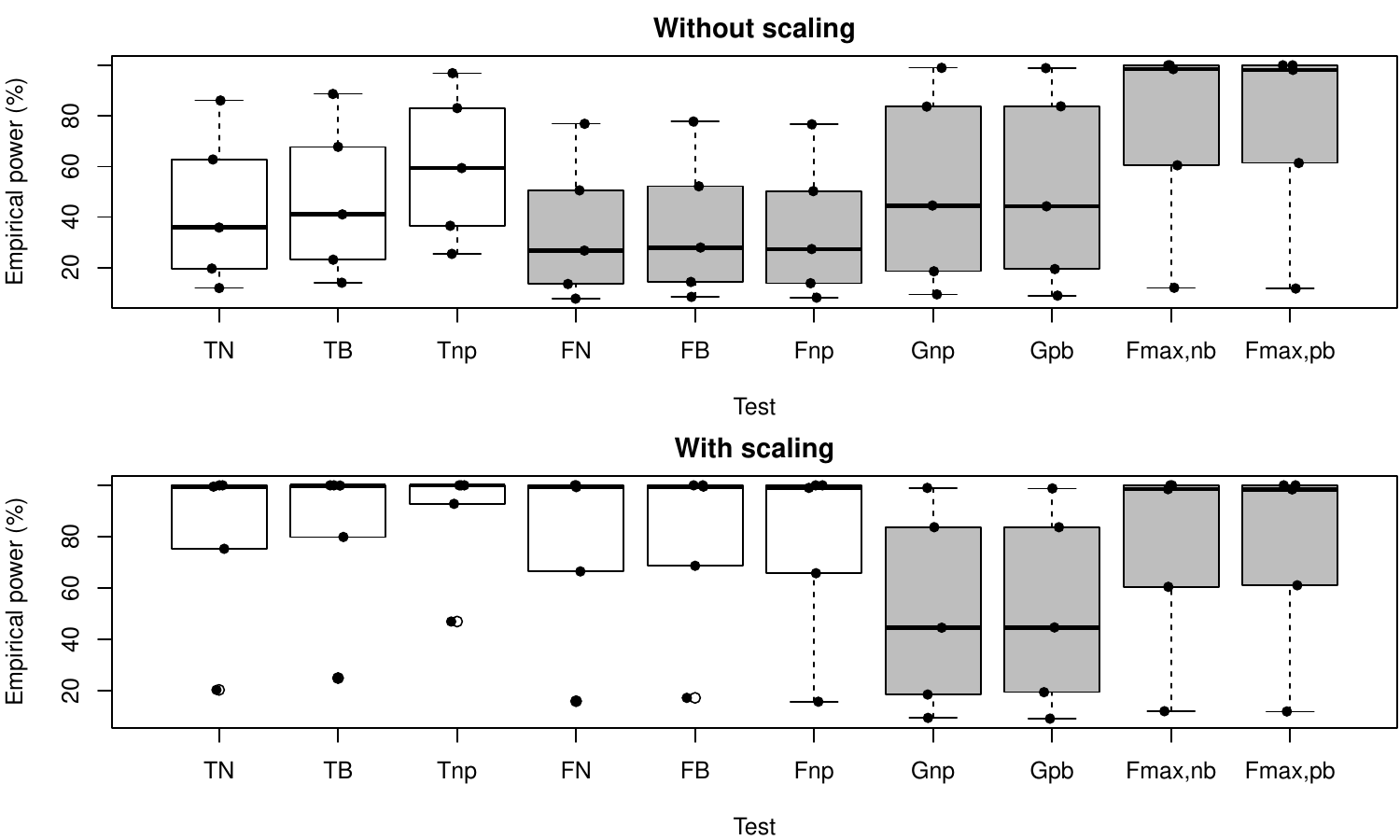}
\caption[Box-and-whisker plots for the empirical powers obtained in Case~3]{Box-and-whisker plots for the empirical powers (as percentages) obtained in Case~3. White boxplots represent too liberal tests.} 
\label{fig_s_4}
\end{figure}

\paragraph{Simulation results} The complete results of the simulation study are presented in Tables~S1-S6 in the supplement. Here, we summarise the results in Figures~\ref{fig_s_1}-\ref{fig_s_4}.

First, we consider the case without scaling (Figures~\ref{fig_s_1},~\ref{fig_s_2}, and \ref{fig_s_4}; Tables~S1-S3 in the supplement). The $L^2$-norm-based tests reveal a tendency towards being highly liberal. The $F_n^B$ and $F_n^{nb}$ may also exhibit an overly liberal character for highly correlated functional data ($\rho=0.1,0.3$), which is not acceptable. On the other hand, the $F_n^N$, $F_n^{nb}$, and both $G_n$-based tests are conservative in cases of less correlation ($\rho=0.9$). Fortunately, for other settings, the new tests $G_n^{nb}$ and $G_n^{pb}$ control the type I error level accurately, which is also true for the $F_{\max,n}$-based tests across all settings. 

Let us now consider the power of the tests. For a fair comparison, we focus only on the non-liberal tests. For highly correlated functional data (Case~3 and $\rho = 0.1,0.3$ in Cases~1–2), the $F_{\max,n}$ tests demonstrate the highest power. However, in cases of lower correlation ($\rho = 0.7,0.9$ in Cases~1–2), the power of the aforementioned tests tends to be lower. In these situations, the $F_n^N$, $F_n^{nb}$ and both bootstrap $G_n$ tests exhibit the highest power. Additionally, the new $G_n$ tests are at least slightly better than the $F_n$-based tests. Moreover, in Case~3, the $G_n$-based tests outperform the $F_n$ tests. At medium correlation levels ($\rho=0.5$ in Cases~1–2), all tests perform similarly in terms of power, but the $F_{\max,n}$ tests can still be more powerful than the other tests. Among the new tests, both bootstrap methods perform very similarly.

Let us now turn to the case with scaling (Figures~\ref{fig_s_1},~\ref{fig_s_3}, and \ref{fig_s_4}; Tables~S4-S6 in the supplement). For the new tests, which are scale-invariant, the results remain consistent with those obtained without scaling. However, for the $T_n$ and $F_n$ tests, we see a strong impact, particularly in terms of power. Specifically, these tests have much less power than before scaling in Cases~1-2. On the other hand, in Case~3, they all are too liberal.

In summary, the new tests are promising as they effectively control the type I error level and have the best power. More precisely, the $F_{\max,n}$-based tests demonstrate the highest power for highly correlated functional data, which is common in practice. On the other hand, the new $G_n$ tests are particularly effective under lower correlation.

\section{Real data example}
\label{sec_5}
We provide tests to verify the statistical significance of the factors in the audible noise data set. Specifically, we test the null hypothesis \eqref{GLHT} for each factor from A to G, setting matrices $\mathbf{C}$ of size $1 \times 15$ as the subsequent rows of matrix~\eqref{mat_c} and $\mathbf{c}(t)=0$. The estimated main-effect contrast functions for factors A-G are presented in Figure~\ref{fig_2}. The $p$-values of the tests are shown in Table~\ref{tab_2} and Table~S7 in the supplement. These tables also provide the empirical sizes and powers of the tests, which were obtained through a simulation study based on this data set for a deeper analysis. The study is described in the supplement.

\begin{figure}[!ht] 
    \centering
    \includegraphics[width=0.99\textwidth,height=0.4\textheight]{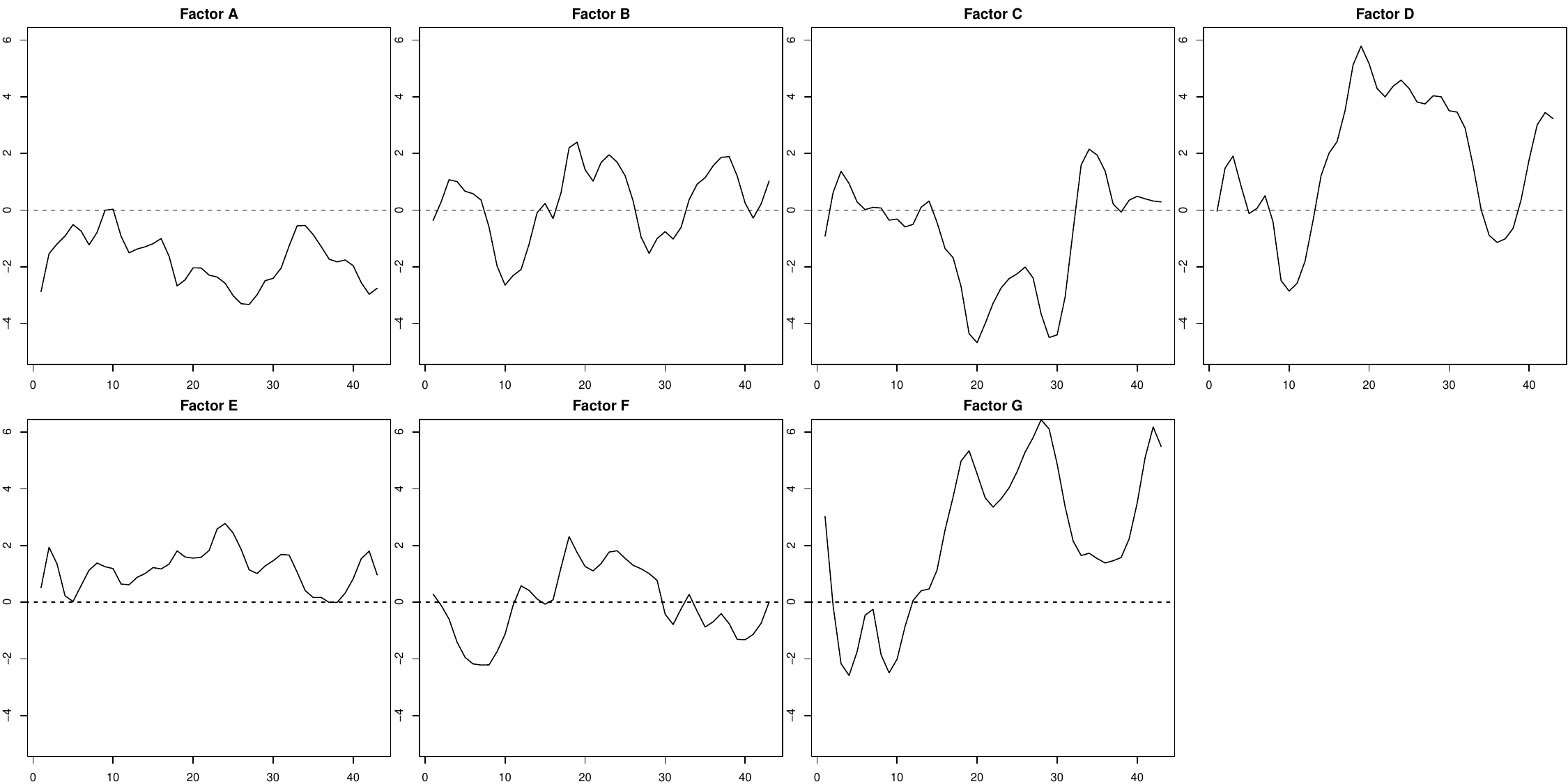}
\caption{Estimated main-effect contrast functions for factors in the audible noise data.}
\label{fig_2}
\end{figure}

Let us discuss the results in Table~\ref{tab_2}. All tests consistently indicate that factors C, D and G are highly significant, while factors B and E are not significant. For factor A, we can also conclude significant differences, although the $F_{\max,n}$-based tests are borderline between rejecting and accepting the null hypothesis. In contrast, the $F_{\max,n}^{nb}$ test rejects the null for factor F, while $F_{\max,n}^{pb}$ test is again borderline, unlike the other tests. This discrepancy can be explained by the fact that the $F_{\max,n}$-based tests outperform the other tests in terms of power. We can also observe that, in general, the new tests are more powerful than the $T_n$ and $F_n$ tests. Some of the latter tests are too liberal compared to the new ones.

For illustrative purposes, we also consider the scale-invariance. As in Section~\ref{sec_4}, we multiply the functions of sound pressure levels by $h(t)=1/(t+1/43)$ and then apply the tests. The results are presented in Table~S7 of the supplement. For the new tests, the results remain consistent with those obtained before scaling. However, the $p$-values, empirical sizes, and powers of the $T_n$ and $F_n$ tests usually change significantly. In particular, their decisions for factors~C and~D are opposite to the previous ones, and their power is much smaller than with the original data.

\begin{table}[!t]
\centering
\captionsetup{list=no}
\begin{tabular}{ccccccccccc}
\hline
  & $T_n^N$ & $T_n^B$  & $T_n^{nb}$ & $F_n^N$ & $F_n^B$  & $F_n^{nb}$  & $G_n^{nb}$ & $G_n^{pb}$ & $F_{\max,n}^{nb}$ & $F_{\max,n}^{pb}$ \\
\hline
\multicolumn{11}{l}{$P$-values}\\
A & \textbf{0.024} & \textbf{0.015} & \textbf{0.011} & 0.029 & \textbf{0.020} & 0.035 & 0.032 & 0.035 & 0.058 & 0.055\\
B & \textbf{0.324} & \textbf{0.314} & \textbf{0.177} & 0.330 & \textbf{0.328} & 0.283 & 0.357 & 0.364 & 0.626 & 0.605\\
C & \textbf{0.011} & \textbf{0.006} & \textbf{0.004} & 0.014 & \textbf{0.008} & 0.014 & 0.037 & 0.035 & 0.036 & 0.032\\
D & \textbf{0.000} & \textbf{0.000} & \textbf{0.000} & 0.000 & \textbf{0.000} & 0.000 & 0.001 & 0.000 & 0.002 & 0.001\\
E & \textbf{0.278} & \textbf{0.265} & \textbf{0.134} & 0.285 & \textbf{0.279} & 0.242 & 0.290 & 0.314 & 0.682 & 0.673\\
F & \textbf{0.408} & \textbf{0.408} & \textbf{0.236} & 0.412 & \textbf{0.420} & 0.349 & 0.172 & 0.208 & 0.041 & 0.051\\
G & \textbf{0.000} & \textbf{0.000} & \textbf{0.000} & 0.000 & \textbf{0.000} & 0.000 & 0.000 & 0.000 & 0.000 & 0.000\\
\multicolumn{11}{l}{Empirical sizes}\\
&\textbf{7.0}&\textbf{8.9}&\textbf{27.6}&5.3&\textbf{6.5}&5.0&4.9&4.8&4.5&5.0\\
\multicolumn{11}{l}{Empirical powers}\\
A&\textbf{71.0}&\textbf{74.3}&\textbf{81.6}&69.6&\textbf{72.5}&69.5&75.8&76.2&74.9&74.9\\
B&\textbf{38.6}&\textbf{45.0}&\textbf{59.9}&35.2&\textbf{40.7}&36.0&40.7&40.8&44.9&43.8\\
C&\textbf{83.6}&\textbf{86.5}&\textbf{93.9}&81.3&\textbf{84.9}&81.8&78.9&78.5&81.0&80.7\\
D&\textbf{97.4}&\textbf{98.0}&\textbf{99.1}&97.2&\textbf{97.7}&97.1&98.0&97.9&96.2&96.4\\
E&\textbf{37.5}&\textbf{43.0}&\textbf{53.6}&35.7&\textbf{40.1}&36.5&40.3&40.7&38.4&39.1\\
F&\textbf{26.9}&\textbf{32.5}&\textbf{49.2}&24.8&\textbf{29.4}&25.6&53.6&53.1&74.3&73.8\\
G&\textbf{99.7}&\textbf{99.9}&\textbf{99.9}&99.4&\textbf{99.7}&99.7&100.0&100.0&100.0&100.0\\
\hline
\end{tabular}
\vspace{10pt}  
\caption{$P$-values of tests verifying significance of all factors and the corresponding empirical sizes and powers (as percentages). Too liberal results are presented in bold, i.e., when the empirical size is greater than $6.4\%$ - the upper limit of $95\%$ binomial confidence interval for empirical size \citePaper{DuchesneFrancq2015}.}
\label{tab_2}
\end{table}

\section{Conclusions}
\label{sec_6}
In this paper, we examined the globalizing pointwise $F$-test and the $F_{\max}$-test for the general linear hypothesis testing in an ill-conditioned functional response model. We successfully applied nonparametric and parametric bootstrap approaches to construct the tests without any distributional assumption. The new test procedures outperformed other tests in terms of size control and statistical power. Moreover, in contrast to the known tests, the new tests are scale-invariant. The power of the $F_{\max}$-tests is found to be higher (respectively lower) than that of the globalizing pointwise $F$-tests when there is a strong (respectively weak) correlation between observations at distinct time points within the functional data.

\paragraph{Supplement} In the supplement, we present the simulation results from Section~\ref{sec_4} and provide details for the audible noise data example discussed in Section~\ref{sec_5}.

\paragraph{Acknowledgments} This research is based on the partial results of the project ``Initiative of Excellence - Research University" funded by grant number 118/34/ID-UB/0027.

{\small
\bibliographystylePaper{unsrt}
\bibliographyPaper{bibliography_LS_NS}}

\clearpage

\setcounter{page}{1}  

\begin{center}
\Large\bfseries Supplementary materials to\\
General linear hypothesis testing\\ in ill-conditioned functional response model
\end{center}
\begin{center}
    Łukasz Smaga, \and Natalia Stefańska\footnote{Corresponding author}
\end{center}

\maketitle

\hrule
\vspace{1cm}


This supplement contains the simulation results of Section~4 and the details for the audible noise data example of Section~5.

\vspace{0.5em}

\setcounter{section}{0}
\renewcommand{\thesection}{S\arabic{section}} 

\setcounter{table}{0} 
\renewcommand{\thetable}{S\arabic{table}} 

\startcontents[sections]
\printcontents[sections]{l}{1}{\setcounter{tocdepth}{3}}

\listoftables

\clearpage

\section{Simulation results of Section 4}
\subsection{Tables}
In Tables~\ref{tab_sr_1}–\ref{tab_sr_6}, we present all the results from the simulation study in Section~4 of the main paper.

\begin{table}[!h]
\centering
\begin{tabular}{ccrrrrrrrrrr}
\hline
 $\rho$ & $\delta$ & $T_n^N$ & $T_n^B$  & $T_n^{nb}$ & $F_n^N$ & $F_n^B$  & $F_n^{nb}$  & $G_n^{nb}$ & $G_n^{pb}$ & $F_{\max,n}^{nb}$ & $F_{\max,n}^{pb}$ \\
\hline
0.1&0.00&\textbf{10.2}&\textbf{12.0}&\textbf{22.2}&6.4&\textbf{6.6}&\textbf{6.7}&6.2&6.3&5.6&6.1\\
&0.02&\textbf{15.9}&\textbf{19.0}&\textbf{29.2}&9.2&\textbf{9.7}&\textbf{8.9}&9.2&10.1&15.9&16.5\\
&0.04&\textbf{32.7}&\textbf{38.4}&\textbf{55.9}&23.7&\textbf{24.2}&\textbf{24.0}&24.3&23.8&58.4&57.6\\
&0.06&\textbf{72.5}&\textbf{77.9}&\textbf{89.2}&57.2&\textbf{58.4}&\textbf{57.0}&58.2&59.5&95.9&95.9\\
&0.08&\textbf{96.2}&\textbf{97.6}&\textbf{99.4}&90.6&\textbf{91.0}&\textbf{90.3}&90.9&91.4&99.9&99.8\\
&0.10&\textbf{99.6}&\textbf{99.7}&\textbf{100.0}&99.1&\textbf{99.1}&\textbf{99.2}&99.2&99.2&100.0&100.0\\\addlinespace
0.3&0.00&\textbf{9.2}&\textbf{11.5}&\textbf{24.4}&6.4&\textbf{6.7}&6.4&6.3&6.3&6.4&6.2\\
&0.04&\textbf{17.2}&\textbf{19.9}&\textbf{34.1}&10.2&\textbf{11.3}&10.7&12.2&11.8&13.7&13.5\\
&0.06&\textbf{26.0}&\textbf{29.1}&\textbf{49.3}&19.4&\textbf{20.9}&19.4&19.1&19.0&28.9&29.0\\
&0.08&\textbf{42.7}&\textbf{48.4}&\textbf{69.1}&31.5&\textbf{33.3}&31.9&33.1&32.9&57.6&56.7\\
&0.10&\textbf{64.3}&\textbf{70.6}&\textbf{86.3}&52.3&\textbf{54.6}&52.1&54.5&55.1&84.5&85.4\\
&0.12&\textbf{84.7}&\textbf{87.4}&\textbf{97.0}&76.3&\textbf{77.5}&75.9&77.4&77.6&95.8&96.2\\\addlinespace
0.5&0.00&\textbf{8.3}&\textbf{10.2}&\textbf{26.1}&4.8&5.9&4.6&5.2&4.8&4.2&4.0\\
&0.05&\textbf{14.5}&\textbf{17.3}&\textbf{38.4}&9.5&10.8&9.3&10.0&10.5&10.4&10.6\\
&0.10&\textbf{37.3}&\textbf{42.2}&\textbf{67.5}&27.8&30.9&28.8&29.4&28.8&33.0&33.3\\
&0.15&\textbf{76.4}&\textbf{79.7}&\textbf{94.1}&66.6&69.8&66.4&67.9&68.9&80.1&80.3\\
&0.20&\textbf{97.2}&\textbf{98.0}&\textbf{99.6}&95.3&96.0&95.1&95.1&95.2&98.3&98.6\\
&0.25&\textbf{99.8}&\textbf{100.0}&\textbf{100.0}&99.6&99.7&99.5&99.5&99.4&100.0&100.0\\\addlinespace
0.7&0.00&\textbf{7.0}&\textbf{10.8}&\textbf{34.5}&4.6&6.1&4.2&5.1&5.2&5.9&5.4\\
&0.10&\textbf{21.9}&\textbf{28.1}&\textbf{59.0}&15.5&19.4&14.9&16.6&16.4&13.3&13.4\\
&0.15&\textbf{47.0}&\textbf{53.8}&\textbf{82.1}&38.7&44.9&38.2&39.0&39.9&30.1&30.9\\
&0.20&\textbf{77.8}&\textbf{83.1}&\textbf{96.4}&70.0&75.2&70.1&71.0&71.6&62.8&63.3\\
&0.25&\textbf{94.8}&\textbf{96.7}&\textbf{99.8}&92.8&93.9&92.3&92.5&92.9&87.8&88.2\\
&0.30&\textbf{99.7}&\textbf{99.8}&\textbf{100.0}&99.2&99.5&99.0&99.2&99.2&97.4&97.2\\\addlinespace
0.9&0.00&4.3&\textbf{7.7}&\textbf{44.8}&2.7&4.8&2.9&3.3&2.9&5.2&5.2\\
&0.20&38.6&\textbf{48.9}&\textbf{84.8}&30.3&40.3&30.3&31.1&31.3&17.3&17.7\\
&0.25&60.4&\textbf{70.0}&\textbf{94.6}&52.3&62.1&51.7&52.3&53.0&31.6&31.0\\
&0.30&80.6&\textbf{87.4}&\textbf{99.1}&73.7&81.6&73.6&73.5&74.2&48.0&48.3\\
&0.35&94.7&\textbf{97.1}&\textbf{99.9}&90.7&95.0&90.9&91.1&90.7&66.6&68.1\\
&0.40&99.2&\textbf{99.6}&\textbf{100.0}&98.1&99.3&98.2&97.4&97.7&83.5&83.0\\
\hline
\end{tabular}
\vspace{10pt}  
\caption[Empirical sizes and powers obtained in Case~1 without scaling]{Empirical sizes and powers (as percentages) for all tests obtained in Case~1 without scaling. Too liberal results are presented in bold. (We define the test as too liberal, when its empirical size is greater than $6.4\%$, which is the value of the upper limit of $95\%$ binomial confidence interval for empirical size for 1000 simulation runs \citeSupp{DuchesneFrancq2015supp}.)}
\label{tab_sr_1}
\end{table}

\begin{table}[t]
\centering
\begin{tabular}{ccrrrrrrrrrr}
\hline
 $\rho$ & $\delta$ & $T_n^N$ & $T_n^B$  & $T_n^{nb}$ & $F_n^N$ & $F_n^B$  & $F_n^{nb}$  & $G_n^{nb}$ & $G_n^{pb}$ & $F_{\max,n}^{nb}$ & $F_{\max,n}^{pb}$ \\
\hline
0.1&0.00&\textbf{8.7}&\textbf{10.8}&\textbf{19.4}&4.7&4.7&4.5&4.5&4.8&4.0&4.2\\
&0.02&\textbf{14.3}&\textbf{18.1}&\textbf{28.1}&9.1&9.4&9.4&9.6&9.3&16.8&16.8\\
&0.04&\textbf{38.5}&\textbf{44.3}&\textbf{58.0}&27.3&28.1&28.0&29.0&28.5&64.0&64.1\\
&0.06&\textbf{73.3}&\textbf{77.7}&\textbf{86.0}&62.5&63.0&64.0&64.4&62.9&93.9&93.5\\
&0.08&\textbf{93.3}&\textbf{94.6}&\textbf{96.2}&88.5&88.6&89.9&89.6&88.3&99.2&99.1\\
&0.10&\textbf{98.0}&\textbf{98.5}&\textbf{98.8}&96.6&96.6&97.1&97.2&96.5&99.9&99.9\\\addlinespace
0.3&0.00&\textbf{7.8}&\textbf{10.3}&\textbf{21.1}&4.2&4.6&4.3&4.9&4.7&4.8&5.0\\
&0.04&\textbf{15.3}&\textbf{18.5}&\textbf{33.1}&10.5&11.3&10.7&11.8&11.2&14.4&13.9\\
&0.06&\textbf{27.9}&\textbf{32.2}&\textbf{50.3}&19.9&20.8&20.3&21.7&21.0&34.7&33.8\\
&0.08&\textbf{47.4}&\textbf{53.2}&\textbf{68.0}&36.9&38.6&37.3&38.7&38.4&60.8&60.7\\
&0.10&\textbf{67.9}&\textbf{72.4}&\textbf{83.9}&58.3&60.0&59.1&60.1&59.2&82.6&82.7\\
&0.12&\textbf{83.9}&\textbf{87.0}&\textbf{92.3}&76.5&76.9&77.0&78.1&77.2&93.6&93.4\\\addlinespace
0.5&0.00&\textbf{7.0}&\textbf{9.4}&\textbf{23.1}&4.6&4.9&4.5&4.9&4.7&5.4&5.1\\
&0.05&\textbf{12.3}&\textbf{15.9}&\textbf{34.0}&8.3&9.5&8.8&10.2&9.9&9.3&9.2\\
&0.10&\textbf{39.9}&\textbf{45.1}&\textbf{66.2}&29.9&33.4&30.0&32.1&31.6&37.5&37.0\\
&0.15&\textbf{76.9}&\textbf{80.9}&\textbf{89.6}&69.2&71.7&69.3&71.8&70.9&79.2&79.0\\
&0.20&\textbf{94.5}&\textbf{95.4}&\textbf{97.3}&91.6&92.5&92.3&92.7&92.2&97.1&96.9\\
&0.25&\textbf{98.7}&\textbf{99.4}&\textbf{99.5}&97.6&98.0&98.0&98.2&98.1&99.7&99.5\\\addlinespace
0.7&0.00&\textbf{6.8}&\textbf{9.4}&\textbf{27.7}&4.2&5.5&3.9&5.4&5.0&5.8&5.8\\
&0.10&\textbf{20.9}&\textbf{26.5}&\textbf{56.2}&13.6&17.7&14.1&15.6&15.4&14.0&13.3\\
&0.15&\textbf{49.9}&\textbf{57.1}&\textbf{79.4}&40.0&46.6&41.0&41.4&40.9&30.9&31.4\\
&0.20&\textbf{77.9}&\textbf{82.6}&\textbf{92.0}&70.8&75.6&71.3&73.2&71.3&62.5&61.7\\
&0.25&\textbf{93.3}&\textbf{94.6}&\textbf{96.8}&89.2&91.6&90.0&91.3&90.2&87.5&86.8\\
&0.30&\textbf{98.0}&\textbf{98.4}&\textbf{99.3}&96.4&96.8&96.8&97.1&96.9&96.6&97.1\\\addlinespace
0.9&0.00&5.6&\textbf{8.9}&\textbf{36.6}&3.2&6.1&3.0&3.7&3.7&6.1&6.1\\
&0.20&37.8&\textbf{49.2}&\textbf{79.8}&27.5&38.0&28.4&30.4&30.0&18.1&18.3\\
&0.25&62.5&\textbf{71.4}&\textbf{90.4}&52.2&63.4&52.8&54.6&53.4&29.7&29.3\\
&0.30&81.4&\textbf{85.9}&\textbf{97.0}&75.3&81.6&75.0&75.5&75.8&47.1&46.3\\
&0.35&91.9&\textbf{95.2}&\textbf{98.3}&89.0&91.7&89.0&90.3&89.3&68.4&68.0\\
&0.40&97.5&\textbf{98.3}&\textbf{99.2}&96.8&97.4&96.8&97.2&96.7&83.7&84.2\\
\hline
\end{tabular}
\vspace{10pt}  
\caption[Empirical sizes and powers obtained in Case~2 without scaling]{Empirical sizes and powers (as percentages) for all tests obtained in Case~2 without scaling. Too liberal results are presented in bold. (We define the test as too liberal, when its empirical size is greater than $6.4\%$, which is the value of the upper limit of $95\%$ binomial confidence interval for empirical size for 1000 simulation runs \citeSupp{DuchesneFrancq2015supp}.)}
\label{tab_sr_2}
\end{table}

\begin{table}[t]
\centering
\begin{tabular}{crrrrrrrrrr}
\hline
$\delta$ & $T_n^N$ & $T_n^B$  & $T_n^{nb}$ & $F_n^N$ & $F_n^B$  & $F_n^{nb}$  & $G_n^{nb}$ & $G_n^{pb}$ & $F_{\max,n}^{nb}$ & $F_{\max,n}^{pb}$ \\
\hline
0.00&\textbf{10.1}&\textbf{12.0}&\textbf{22.6}&5.9&6.3&6.0&6.1&5.9&4.6&4.8\\
0.02&\textbf{12.0}&\textbf{14.1}&\textbf{25.5}&7.8&8.5&8.2&9.5&9.0&12.1&11.8\\
0.04&\textbf{19.7}&\textbf{23.2}&\textbf{36.6}&13.6&14.4&13.9&18.6&19.5&60.5&61.4\\
0.06&\textbf{35.9}&\textbf{41.1}&\textbf{59.4}&26.8&28.0&27.4&44.6&44.3&98.5&98.2\\
0.08&\textbf{62.8}&\textbf{67.8}&\textbf{83.1}&50.6&52.2&50.3&83.7&83.8&100.0&100.0\\
0.10&\textbf{86.1}&\textbf{88.7}&\textbf{96.9}&76.9&77.8&76.7&99.0&98.9&100.0&100.0\\
\hline
\end{tabular}
\vspace{10pt}  
\caption[Empirical sizes and powers obtained in Case~3 without scaling]{Empirical sizes and powers (as percentages) for all tests obtained in Case~3 without scaling. Too liberal results are presented in bold. (We define the test as too liberal, when its empirical size is greater than $6.4\%$, which is the value of the upper limit of $95\%$ binomial confidence interval for empirical size for 1000 simulation runs \citeSupp{DuchesneFrancq2015supp}.)}
\label{tab_sr_3}
\end{table}

\begin{table}[t]
\centering
\begin{tabular}{ccrrrrrrrrrr}
\hline
 $\rho$ & $\delta$ & $T_n^N$ & $T_n^B$  & $T_n^{nb}$ & $F_n^N$ & $F_n^B$  & $F_n^{nb}$  & $G_n^{nb}$ & $G_n^{pb}$ & $F_{\max,n}^{nb}$ & $F_{\max,n}^{pb}$ \\
\hline
0.1&0.00&\textbf{10.2}&\textbf{12.7}&\textbf{21.5}&5.6&5.7&5.9&6.4&6.4&5.6&6.1\\
&0.02&\textbf{12.5}&\textbf{15.3}&\textbf{24.6}&7.3&7.3&7.3&9.2&10.1&15.9&16.1\\
&0.04&\textbf{18.8}&\textbf{23.1}&\textbf{35.2}&12.5&12.5&12.3&24.3&23.8&58.4&57.9\\
&0.06&\textbf{34.3}&\textbf{39.1}&\textbf{54.9}&24.2&24.2&24.0&58.2&59.8&95.9&96.0\\
&0.08&\textbf{58.3}&\textbf{64.8}&\textbf{79.7}&43.7&43.7&43.8&90.9&91.2&99.9&99.9\\
&0.10&\textbf{81.8}&\textbf{86.8}&\textbf{95.3}&70.5&70.7&71.4&99.2&99.2&100.0&100.0\\\addlinespace
0.3&0.00&\textbf{9.7}&\textbf{12.6}&\textbf{20.9}&5.2&5.5&5.5&6.3&6.2&6.4&6.3\\
&0.04&\textbf{12.5}&\textbf{15.5}&\textbf{25.5}&6.9&7.1&6.9&12.2&11.6&13.7&13.7\\
&0.06&\textbf{16.1}&\textbf{19.1}&\textbf{30.6}&9.6&9.8&9.8&19.1&19.6&28.9&29.4\\
&0.08&\textbf{20.8}&\textbf{24.8}&\textbf{39.0}&14.2&14.5&14.1&33.1&32.8&57.6&57.0\\
&0.10&\textbf{28.6}&\textbf{33.8}&\textbf{50.1}&19.5&19.8&20.3&54.5&55.3&84.5&84.9\\
&0.12&\textbf{39.6}&\textbf{45.6}&\textbf{63.2}&28.1&28.5&27.5&77.4&77.7&95.8&96.1\\\addlinespace
0.5&0.00&\textbf{9.8}&\textbf{11.2}&\textbf{20.9}&5.1&5.3&5.5&5.2&4.9&4.2&3.8\\
&0.05&\textbf{10.9}&\textbf{13.7}&\textbf{25.0}&6.5&6.6&6.5&10.0&10.0&10.4&9.8\\
&0.10&\textbf{17.7}&\textbf{20.7}&\textbf{34.6}&10.3&10.7&10.3&29.4&29.4&33.0&33.3\\
&0.15&\textbf{30.7}&\textbf{36.0}&\textbf{51.6}&19.9&20.4&19.2&67.9&69.1&80.1&80.6\\
&0.20&\textbf{49.7}&\textbf{55.6}&\textbf{71.8}&38.3&38.9&38.4&95.1&95.0&98.3&98.3\\
&0.25&\textbf{73.1}&\textbf{77.9}&\textbf{89.3}&60.2&61.0&60.2&99.5&99.5&100.0&100.0\\\addlinespace
0.7&0.00&\textbf{9.8}&\textbf{11.5}&\textbf{22.5}&5.0&5.4&5.2&5.1&5.2&5.9&5.5\\
&0.10&\textbf{12.4}&\textbf{14.9}&\textbf{27.6}&7.1&7.7&7.4&16.6&16.6&13.3&13.5\\
&0.15&\textbf{17.7}&\textbf{20.7}&\textbf{35.1}&10.8&11.2&10.9&39.0&39.9&30.1&30.9\\
&0.20&\textbf{24.7}&\textbf{29.2}&\textbf{46.3}&16.4&17.5&16.7&71.0&71.8&62.8&62.6\\
&0.25&\textbf{35.9}&\textbf{40.9}&\textbf{59.2}&25.8&26.9&25.7&92.5&93.1&87.8&87.3\\
&0.30&\textbf{50.5}&\textbf{55.3}&\textbf{71.1}&37.8&38.7&38.0&99.2&99.3&97.4&97.6\\\addlinespace
0.9&0.00&\textbf{8.8}&\textbf{10.8}&\textbf{22.6}&5.0&5.8&5.4&3.3&2.9&5.2&5.1\\
&0.20&\textbf{14.7}&\textbf{16.1}&\textbf{29.7}&8.4&9.1&8.7&31.1&31.0&17.3&17.4\\
&0.25&\textbf{16.3}&\textbf{20.0}&\textbf{35.1}&10.9&12.0&11.0&52.3&53.2&31.6&31.2\\
&0.30&\textbf{20.8}&\textbf{25.3}&\textbf{40.8}&14.5&15.3&14.5&73.5&73.6&48.0&48.5\\
&0.35&\textbf{26.9}&\textbf{29.0}&\textbf{47.9}&18.6&19.5&18.8&91.1&91.0&66.6&68.0\\
&0.40&\textbf{30.7}&\textbf{36.8}&\textbf{54.3}&23.0&25.4&23.3&97.4&97.4&83.5&83.5\\
\hline
\end{tabular}
\vspace{10pt}  
\caption[Empirical sizes and powers obtained in Case~1 with scaling]{Empirical sizes and powers (as percentages) for all tests obtained in Case~1 with scaling. Too liberal results are presented in bold. (We define the test as too liberal, when its empirical size is greater than $6.4\%$, which is the value of the upper limit of $95\%$ binomial confidence interval for empirical size for 1000 simulation runs \citeSupp{DuchesneFrancq2015supp}.)}
\label{tab_sr_4}
\end{table}

\begin{table}[t]
\centering
\begin{tabular}{ccrrrrrrrrrr}
\hline
 $\rho$ & $\delta$ & $T_n^N$ & $T_n^B$  & $T_n^{nb}$ & $F_n^N$ & $F_n^B$  & $F_n^{nb}$  & $G_n^{nb}$ & $G_n^{pb}$ & $F_{\max,n}^{nb}$ & $F_{\max,n}^{pb}$ \\
\hline
0.1&0.00&\textbf{9.0}&\textbf{11.4}&\textbf{19.3}&3.9&3.9&4.6&4.5&5.0&4.0&4.3\\
&0.02&\textbf{11.6}&\textbf{14.8}&\textbf{22.6}&5.9&6.0&6.6&9.6&9.4&16.8&16.7\\
&0.04&\textbf{19.7}&\textbf{24.0}&\textbf{36.0}&12.6&12.7&13.5&29.0&28.4&64.0&64.0\\
&0.06&\textbf{38.8}&\textbf{44.6}&\textbf{58.2}&27.5&27.6&27.9&64.4&63.4&93.9&93.1\\
&0.08&\textbf{63.5}&\textbf{67.4}&\textbf{78.1}&51.0&51.2&52.1&89.6&88.4&99.2&99.1\\
&0.10&\textbf{83.0}&\textbf{86.0}&\textbf{91.1}&73.9&74.0&74.0&97.2&96.1&99.9&99.9\\\addlinespace
0.3&0.00&\textbf{8.6}&\textbf{10.3}&\textbf{19.7}&4.7&4.9&4.8&4.9&4.6&4.8&4.9\\
&0.04&\textbf{11.3}&\textbf{13.8}&\textbf{24.9}&6.6&6.9&6.8&11.8&10.9&14.4&13.9\\
&0.06&\textbf{15.9}&\textbf{19.0}&\textbf{31.5}&9.4&9.5&9.7&21.7&21.1&34.7&33.2\\
&0.08&\textbf{22.5}&\textbf{27.2}&\textbf{39.2}&14.2&14.7&14.5&38.7&38.6&60.8&61.3\\
&0.10&\textbf{32.4}&\textbf{36.4}&\textbf{51.3}&21.9&22.0&22.6&60.1&59.4&82.6&82.8\\
&0.12&\textbf{44.5}&\textbf{49.4}&\textbf{63.6}&31.6&31.6&32.1&78.1&76.9&93.6&93.5\\\addlinespace
0.5&0.00&\textbf{8.8}&\textbf{11.2}&\textbf{19.7}&5.3&5.5&5.2&4.9&4.6&5.4&5.3\\
&0.05&\textbf{10.5}&\textbf{13.2}&\textbf{23.2}&6.5&6.6&6.6&10.2&10.0&9.3&9.5\\
&0.10&\textbf{17.1}&\textbf{21.7}&\textbf{34.4}&11.1&11.2&11.5&32.1&31.6&37.5&37.1\\
&0.15&\textbf{32.0}&\textbf{36.4}&\textbf{50.8}&21.7&22.2&22.4&71.8&70.6&79.2&79.1\\
&0.20&\textbf{51.9}&\textbf{56.8}&\textbf{71.9}&38.0&38.6&39.5&92.7&92.9&97.1&96.9\\
&0.25&\textbf{74.9}&\textbf{79.2}&\textbf{87.7}&62.8&63.8&62.5&98.2&98.1&99.7&99.6\\\addlinespace
0.7&0.00&\textbf{9.6}&\textbf{12.7}&\textbf{21.0}&4.6&4.9&4.9&5.4&5.0&5.8&5.8\\
&0.10&\textbf{13.2}&\textbf{15.2}&\textbf{26.6}&8.1&8.5&8.1&15.6&15.9&14.0&13.1\\
&0.15&\textbf{17.3}&\textbf{21.1}&\textbf{32.6}&12.2&13.0&12.5&41.4&41.9&30.9&31.1\\
&0.20&\textbf{25.1}&\textbf{29.5}&\textbf{44.7}&18.0&18.9&17.9&73.2&72.0&62.5&62.3\\
&0.25&\textbf{35.6}&\textbf{41.4}&\textbf{57.0}&26.6&27.2&26.7&91.3&90.2&87.5&86.4\\
&0.30&\textbf{49.7}&\textbf{55.4}&\textbf{70.8}&38.1&39.6&39.3&97.1&96.8&96.6&96.8\\\addlinespace
0.9&0.00&\textbf{9.5}&\textbf{12.0}&\textbf{20.7}&5.4&5.9&5.4&3.7&3.6&6.1&6.0\\
&0.20&\textbf{13.9}&\textbf{16.3}&\textbf{28.3}&8.7&9.5&8.6&30.4&29.6&18.1&18.1\\
&0.25&\textbf{17.3}&\textbf{20.0}&\textbf{33.5}&11.0&12.0&11.0&54.6&53.6&29.7&30.0\\
&0.30&\textbf{21.3}&\textbf{23.2}&\textbf{39.5}&14.3&14.9&14.4&75.5&76.1&47.1&46.5\\
&0.35&\textbf{24.5}&\textbf{28.3}&\textbf{46.0}&18.7&19.7&18.6&90.3&89.3&68.4&66.9\\
&0.40&\textbf{30.8}&\textbf{37.0}&\textbf{52.8}&22.5&23.9&23.2&97.2&96.9&83.7&83.7\\
\hline
\end{tabular}
\vspace{10pt}  
\caption[Empirical sizes and powers obtained in Case~2 with scaling]{Empirical sizes and powers (as percentages) for all tests obtained in Case~2 with scaling. Too liberal results are presented in bold. (We define the test as too liberal, when its empirical size is greater than $6.4\%$, which is the value of the upper limit of $95\%$ binomial confidence interval for empirical size for 1000 simulation runs \citeSupp{DuchesneFrancq2015supp}.)}
\label{tab_sr_5}
\end{table}

\begin{table}[t]
\centering
\begin{tabular}{crrrrrrrrrr}
\hline
$\delta$ & $T_n^N$ & $T_n^B$  & $T_n^{nb}$ & $F_n^N$ & $F_n^B$  & $F_n^{nb}$  & $G_n^{nb}$ & $G_n^{pb}$ & $F_{\max,n}^{nb}$ & $F_{\max,n}^{pb}$ \\
\hline
0.00&\textbf{8.9}&\textbf{10.7}&\textbf{23.7}&\textbf{6.7}&\textbf{6.8}&\textbf{6.5}&6.1&6.1&4.6&4.6\\
0.02&\textbf{20.4}&\textbf{25.0}&\textbf{47.0}&\textbf{16.0}&\textbf{17.3}&\textbf{15.8}&9.5&9.2&12.1&12.0\\
0.04&\textbf{75.3}&\textbf{79.9}&\textbf{92.8}&\textbf{66.5}&\textbf{68.7}&\textbf{65.8}&18.6&19.5&60.5&61.1\\
0.06&\textbf{99.5}&\textbf{99.9}&\textbf{100.0}&\textbf{99.3}&\textbf{99.5}&\textbf{99.0}&44.6&44.7&98.5&98.4\\
0.08&\textbf{100.0}&\textbf{100.0}&\textbf{100.0}&\textbf{100.0}&\textbf{100.0}&\textbf{100.0}&83.7&83.7&100.0&100.0\\
0.10&\textbf{100.0}&\textbf{100.0}&\textbf{100.0}&\textbf{100.0}&\textbf{100.0}&\textbf{100.0}&99.0&98.8&100.0&100.0\\
\hline
\end{tabular}
\vspace{10pt}  
\caption[Empirical sizes and powers obtained in Case~3 with scaling]{Empirical sizes and powers (as percentages) for all tests obtained in Case~3 with scaling. Too liberal results are presented in bold. (We define the test as too liberal, when its empirical size is greater than $6.4\%$, which is the value of the upper limit of $95\%$ binomial confidence interval for empirical size for 1000 simulation runs \citeSupp{DuchesneFrancq2015supp}.)}
\label{tab_sr_6}
\end{table}

\clearpage

\section{More details for the audible noise data example of Section 5}
In this section, we provide details for the real data example from Section~5 of the main paper.

\subsection{Design matrix}
In the audible noise data, the design matrix is given in~\eqref{mac_x}.

\begin{equation}
\label{mac_x}
\mathbf{X}=\left(\begin{array}{rrrrrrrrrrrrrrr}
1&1&0&1&0&1&0&1&0&1&0&0&1&0&1\\
1&1&0&1&0&1&0&1&0&0&1&0&1&1&0\\
1&1&0&1&0&1&0&0&1&1&0&1&0&1&0\\
1&1&0&1&0&1&0&0&1&0&1&1&0&0&1\\
1&1&0&1&0&0&1&1&0&1&0&1&0&0&1\\
1&1&0&1&0&0&1&1&0&0&1&1&0&1&0\\
1&1&0&1&0&0&1&0&1&1&0&0&1&1&0\\
1&1&0&1&0&0&1&0&1&0&1&0&1&0&1\\
1&1&0&0&1&1&0&1&0&1&0&1&0&1&0\\
1&1&0&0&1&1&0&1&0&0&1&1&0&0&1\\
1&1&0&0&1&1&0&0&1&1&0&0&1&0&1\\
1&1&0&0&1&1&0&0&1&0&1&0&1&1&0\\
1&1&0&0&1&0&1&1&0&1&0&0&1&1&0\\
1&1&0&0&1&0&1&1&0&0&1&0&1&0&1\\
1&1&0&0&1&0&1&0&1&1&0&1&0&0&1\\
1&1&0&0&1&0&1&0&1&0&1&1&0&1&0\\
1&0&1&1&0&1&0&1&0&1&0&1&0&1&0\\
1&0&1&1&0&1&0&1&0&0&1&1&0&0&1\\
1&0&1&1&0&1&0&0&1&1&0&0&1&0&1\\
1&0&1&1&0&1&0&0&1&0&1&0&1&1&0\\
1&0&1&1&0&0&1&1&0&1&0&0&1&1&0\\
1&0&1&1&0&0&1&1&0&0&1&0&1&0&1\\
1&0&1&1&0&0&1&0&1&1&0&1&0&0&1\\
1&0&1&1&0&0&1&0&1&0&1&1&0&1&0\\
1&0&1&0&1&1&0&1&0&1&0&0&1&0&1\\
1&0&1&0&1&1&0&1&0&0&1&0&1&1&0\\
1&0&1&0&1&1&0&0&1&1&0&1&0&1&0\\
1&0&1&0&1&1&0&0&1&0&1&1&0&0&1\\
1&0&1&0&1&0&1&1&0&1&0&1&0&0&1\\
1&0&1&0&1&0&1&1&0&0&1&1&0&1&0\\
1&0&1&0&1&0&1&0&1&1&0&0&1&1&0\\
1&0&1&0&1&0&1&0&1&0&1&0&1&0&1\\
1&0&1&0&1&0&1&0&1&0&1&0&1&0&1\\
1&0&1&0&1&0&1&0&1&0&1&0&1&0&1\\
1&0&1&0&1&0&1&0&1&0&1&0&1&0&1\\
1&0&1&0&1&0&1&0&1&0&1&0&1&0&1\\
\end{array}\right)
\end{equation}

\newpage

\subsection{Simulation study based on the audible noise data example}
In this section, we present the details of the simulation study based on the audible noise dataset. This simulation study differs from the one in Section 4. In Section 4, we investigated the general properties of the tests in more detail, considering, for example, different correlations. In contrast, the new simulation study fits the data set more closely to explain the results of applying the tests to that data set.

Let us describe the simulation study based on the audible noise data example. To mimic the data given in the data set, we generated the simulation data from the multivariate normal distribution with the following specifications:

\begin{itemize}
\item the sample size from the data example, i.e., $n=36$,
\item the design matrix $\mathbf{X}$ as given in~\eqref{mac_x} was used,
\item the covariance matrix for each observation was equal to the sample covariance function of the functional responses,
\item for checking the type I error control: $\boldsymbol{\beta}(t)=\mathbf{0}$,
\item for power investigation: $\boldsymbol{\beta}(t)=\hat{\boldsymbol{\beta}}(t)$.
\end{itemize}

The results of this simulation study are presented in Table~1 in the main paper and Table~\ref{tab_3} for the scaled data.

\begin{table}[!t]
\centering
\begin{tabular}{ccccccccccc}
\hline
  & $T_n^N$ & $T_n^B$  & $T_n^{nb}$ & $F_n^N$ & $F_n^B$  & $F_n^{nb}$  & $G_n^{nb}$ & $G_n^{pb}$ & $F_{\max,n}^{nb}$ & $F_{\max,n}^{pb}$ \\
\hline
\multicolumn{11}{l}{$P$-values}\\
A&\textbf{0.050}&\textbf{0.042}&\textbf{0.017}&0.058&0.056&0.051&0.032&0.039&0.056&0.054\\
B&\textbf{0.705}&\textbf{0.705}&\textbf{0.715}&0.707&0.716&0.798&0.368&0.366&0.611&0.606\\
C&\textbf{0.509}&\textbf{0.503}&\textbf{0.415}&0.514&0.520&0.500&0.046&0.045&0.043&0.038\\
D&\textbf{0.448}&\textbf{0.439}&\textbf{0.331}&0.454&0.459&0.430&0.000&0.000&0.005&0.004\\
E&\textbf{0.494}&\textbf{0.486}&\textbf{0.418}&0.498&0.504&0.502&0.289&0.306&0.695&0.695\\
F&\textbf{0.658}&\textbf{0.656}&\textbf{0.621}&0.661&0.669&0.714&0.206&0.192&0.047&0.054\\
G&\textbf{0.030}&\textbf{0.024}&\textbf{0.019}&0.037&0.040&0.044&0.000&0.000&0.000&0.000\\
\multicolumn{11}{l}{Empirical sizes}\\
&\textbf{8.1}&\textbf{9.8}&\textbf{19.9}&4.7&5.1&5.1&4.9&5.3&4.5&4.9\\
\multicolumn{11}{l}{Empirical powers}\\
A&\textbf{51.2}&\textbf{53.0}&\textbf{61.1}&48.4&49.0&48.5&74.7&75.2&72.7&72.7\\
B&\textbf{7.7}&\textbf{8.7}&\textbf{12.3}&6.7&6.9&6.8&35.2&36.6&40.7&41.2\\
C&\textbf{11.0}&\textbf{12.1}&\textbf{16.4}&9.6&9.8&9.6&77.3&78.7&80.4&80.2\\
D&\textbf{11.2}&\textbf{12.5}&\textbf{19.5}&9.7&10.1&10.1&97.8&97.6&96.8&96.5\\
E&\textbf{12.0}&\textbf{13.8}&\textbf{18.0}&10.3&10.6&10.5&38.1&37.8&37.4&37.9\\
F&\textbf{6.5}&\textbf{7.8}&\textbf{11.8}&5.3&5.4&6.1&56.2&55.8&70.3&70.1\\
G&\textbf{64.3}&\textbf{67.2}&\textbf{76.2}&61.1&61.8&62.4&99.8&99.8&99.9&99.9\\
\hline
\end{tabular}
\vspace{10pt}  
\caption[$P$-values of tests verifying significance of all factors and the corresponding empirical sizes and powers with scaling]{$P$-values of tests verifying significance of all factors and the corresponding empirical sizes and powers (as percentages) with scaling. Too liberal results are presented in bold, i.e., when the empirical size is greater than $6.4\%$ - the upper limit of $95\%$ binomial confidence interval for empirical size for 1000 simulation runs \citeSupp{DuchesneFrancq2015supp}.}
\label{tab_3}
\end{table}

\setcounter{enumiv}{0}

\bibliographystyleSupp{unsrt}
\bibliographySupp{bibliography_LS_NS}

\end{document}